\documentclass{ifacconf}
\usepackage{graphicx}      % include this line if your document contains figures
\usepackage{natbib}        % required for bibliography
\usepackage{xcolor}
\usepackage{amssymb}
\usepackage[ruled]{algorithm2e}  % comment line 670 and 671 in ifacconf.cls (definition alghorithm) %alternativ:boxed 
\usepackage{enumitem}
\usepackage{url}
\usepackage{xspace}
\usepackage{soul}
\usepackage{mathtools}
\usepackage{makecell}
\usepackage{nicefrac}
\usepackage{float}
\usepackage{wrapfig}
\newcommand{\ra}{\rightarrow}

\newtheorem{dfn}{Definition}
\newcommand{\R}{\ensuremath{\mathbb{R}}}
\newcommand{\C}{\ensuremath{\mathbb{C}}}
\newcommand{\bs}[1]{\ensuremath{\boldsymbol{#1}}}

\DeclareMathOperator*{\argmin}{arg\,min}
\DeclareMathOperator*{\argmax}{arg\,max}

\newcommand{\reffig}[1]{Fig.\,\ref{#1}}  
\newcommand{\blue}[1]{\textcolor{blue}{#1}}
\definecolor{dgreen}{rgb}{0.0, 0.5, 0.0}

\newcommand{\U}[1]{\mathrm{#1}}	% Logical formula separator

\makeatletter
\newcommand{\subalign}[1]{%
	\vcenter{%
		\Let@ \restore@math@cr \default@tag
		\baselineskip\fontdimen10 \scriptfont\tw@
		\advance\baselineskip\fontdimen12 \scriptfont\tw@
		\lineskip\thr@@\fontdimen8 \scriptfont\thr@@
		\lineskiplimit\lineskip
		\ialign{\hfil$\m@th\scriptstyle##$&$\m@th\scriptstyle{}##$\crcr
			#1\crcr
		}%
	}
}

\begin{document}
\begin{frontmatter}

\title{Prototypical Description and Controller Design for a Set of Systems Using $\nu$-gap Based Clustering }

\author{Lukas Munser\;\;} 
\author{Arne-Jens Hempel\;\;}
\author{Grigory Devadze\;\;} 
\author{Stefan Streif}

\address{Technische Universit{\"a}t Chemnitz, Automatic Control and System Dynamics Lab, Germany (e-mail: \{lukas.munser,arne-jens.hempel, grigory.devadze,stefan.streif\}@etit.tu-chemnitz.de)}

\begin{abstract}
We present an approach to design stabilizing controllers for a set of linear systems without restrictions regarding their modeling order.
To this end, the systems are treated as abstract objects in the space of the $\nu$-gap metric. 
Via a cluster analysis the set of systems is split into $\nu$-gap similar clusters which are treated separately. 
For this purpose we provide an algorithm that constructs an explicit prototype system by generalizing the information of a given set of systems. 
Applying this algorithm to each cluster a set of prototype systems is obtained.
Given these prototypes we design controllers in such a way that all systems assigned to a cluster will be stabilized by a corresponding controller.
The approach is demonstrated for a set of 80 linear systems.
%\blue{algorithm for SISO only, complexity and convergence and on the relaxations that they require}
\end{abstract}

\begin{keyword}
Robust control, Classification, Model-based control, Distance transformation, Stabilizing controllers
% Five to ten keywords, preferably chosen from the IFAC keyword list.
\end{keyword}

\end{frontmatter}

%===============================================================================
% section Introduction
%===============================================================================
\section{Introduction}
\vspace{-2mm}
In this paper we address the task of automatic controller design for sets of linear time invariant multi-input multi-output (LTI MIMO) systems from a machine learning point of view.
Therefor dynamical systems are perceived as abstract objects in some space.
Together with a suitable metric, similarities and dissimilarities between these objects/systems can be explored paving the way to group similar systems together by techniques of cluster analysis.
The representation of a cluster---of similar systems---by a single but more general prototype system forms the foundation for the design of stabilizing controllers for the respective cluster.
As a result the task of control design for sets of systems can approached systematically in terms of  clustering or classification.
\\
More formally we are concerned with the following problem statement.
\subsubsection{Problem Statement.}
Consider the following set of $n$ LTI MIMO control systems represented in the space of real valued transfer function matrices $\mathcal{R}$:
\begin{equation*}
	\mathcal{G}= \{\bs{G}_i(s)\}, \bs{G}_i(s) \in \mathcal{R}^{p \times q}, p,q\in \mathbb{N} ,\forall i=1,\ldots, n.
\end{equation*}%
For $\mathcal{G}$ find a partition $\mathcal{C} = \{\mathcal{G}_1,\ldots,\mathcal{G}_m \}$, called cluster configuration, such that the systems $\bs{G}_i(s)$ with similar closed loop behavior are grouped into clusters $\mathcal{G}_j$.
For each cluster $\mathcal{G}_j \in \mathcal{C}$ construct a prototype system, called class description, $\bs{G}_{\text{proto},j}(s)$ together with its corresponding controller $\bs{G}_{\text{c},j}(s)$
such that each $\bs{G}_i \in \mathcal{G}_j$ will be stabilized, i.e.
\begin{equation}
	\begin{split}
		&\dfrac{1}{s} \left(\bs{I} + \bs{G}_{\text{c},j}(s)\bs{G}_{i}(s)\right)^{-1} \bs{G}_{\text{c},j}(s)\bs{G}_{i}(s) \leq \infty \\
		&\forall i=1,\ldots,n ,\forall j=1,\ldots,m \text{ and } \forall \bs{G}_{i}(s) \in \mathcal{G}_j.
	\end{split}
\end{equation}
As a consequence every real part of the poles of the closed loop systems have to be less or equal than zero.

%\begin{equation}
%	\begin{split}
%		&\lim_{s\ra0} \left(\bs{I} + \bs{G}_{\text{c},j}(s)\bs{G}_{i}(s)\right)^{-1} \bs{G}_{\text{c},j}(s)\bs{G}_{i}(s)=\bs{0} \\
%		&\forall i=1,\ldots,n ,\forall j=1,\ldots,m \text{ and } \forall \bs{G}_{i}(s) \in \mathcal{G}_j.
%	\end{split}
%\end{equation}

Such a treatment for set of systems can be of interest in many areas of control theory for example in Multi-Model-Adaptive-Control-schemes (MMAC) and for the control of ensemble or hierarchical systems.
The main idea of MMAC is the use of a set of candidate plant models each with an associated controller and a suitable switching procedure \cite[]{baldi2012,tan2017}.
A major question in MMAC is the selection of prototypical systems and models thereof \cite[]{anderson2000,french2008}.
\cite{kersting2018} presented an approach to distribute such prototype models over a given uncertainty set handling models of the same structure only.
\\
Similar challenges occur for the control of ensemble systems.
That is the control of sets of (physically) similar systems, e.g. wind farms, micro grids and many more, with the same control goal \cite[]{Morton:2007,Ali:2013,Petzke:2018}.
\\
Depending on the specific areas or tasks, control systems can be assessed by different metrics.
Considering metrics for dynamical systems some work has been done to compare systems by the help of their open-loop (state)trajectories.
\cite{smola2004} introduced the so called Binet-Cauchy Kernel for the comparison of video sequences via their embedding in linear time-invariant systems.
\\
Regarding the task of controller design the $\nu$-gap metric, introduced by \cite{vinnicombe1993,reinelt2001}, has the capabilities to assess closed-loop systems behavior.
It was introduced in the field of robust control to measures the maximum distance between a nominal and a perturbed plant for which the perturbed plant can be stabilized by the controller designed for the nominal plant.\\
In this context the cluster analysis---of the set of dynamic systems to be controlled---reveals groups of $\nu$-gap-similar systems, which can be controlled by a certain controller.
Accordingly the actual controller design can be performed robustly \cite[]{skogestad2001, mcfarlane1992} based on a prototype system for each the group of systems, accounting for structured uncertainties \cite[]{kersting2018}.

\subsubsection{Contribution and outline of this work.} 
A main contribution of this paper is the derivation of a procedure to generate prototypes for a set of LTI MIMO systems. 
This procedure is not limited by a given model structure/order. 
We also present an approach to find clusters of similar LTI MIMO systems and their corresponding stabilizing controllers, see the following structure of the paper. 

In Section \ref{sec:Distance_metric} a control-oriented metric for closed loop systems is introduced.
Since this metric is solely derived on closed loop systems it directly provides stabilizing controllers.
Based upon this metric, the cluster analysis of dynamical systems as well as the construction of prototypical systems, including an algorithm, will be covered in Section \ref{sec:sys_aggregation}. 
The algorithm is implemented for single-input single output (SISO) systems only, but can be extended for MIMO cases within the framework of the prior theory.
The paper is round off with a case study summarizing the entire approach in Section \ref{sec:CaseStudy}.
Here we also introduce a way to visualize the dissimilarities between systems  in a two dimensional space using a technique called ``t-SNE''.

%===============================================================================
%  section A Distance Metric for Dynamic Systems
%===============================================================================
\section{A Distance Metric for Closed Loop Systems: The $\nu$-Gap}\label{sec:Distance_metric}
Partitioning the set $\mathcal{G}$ by a cluster analysis requires a suitable metric.
Suitable in this context means that the metric measures the closed loop behavior of the systems for all possible combinations of bounded input and output signals.
A promising candidate metric which advances the concept of BIBO-stability for closed loop systems with respect to robustness was introduced by \cite{vinnicombe1993} as the \emph{$\nu$-gap metric}.
\\
For the sake of readability, we will drop the argument $s$ of the transfer functions in the remainder.

%===============================================================================
% subsection nu-gap Metric
%===============================================================================
\subsection{The $\nu$-Gap-Distance Between Two Systems}
%===============================================================================
% subsection Coprime Factor Representations
%===============================================================================
To introduce and to actually calculate the $\nu$-gap distance between two systems their normalized coprime factorizations are required \cite[]{koenings2016}. 
\begin{dfn}[NRCF]
Let $\mathcal{RH}_\infty$ be the space of all proper real transfer functions with stable poles two matrices $\bs{M} \in \mathcal{RH}_\infty^{q \times q}$ and $\bs{N} \in \mathcal{RH}_\infty^{p \times q}$ form a normalized right coprime factorization (NRCF)
\begin{equation}
	\bs{G}(s) = \bs{N}\bs{M}^{-1},
\end{equation}
if $\exists \bs{X}  \in \mathcal{RH}_\infty^{q \times q},\bs{Y}  \in \mathcal{RH}_\infty^{q \times p}$ such that
\begin{equation}
	\begin{split}
		\bs{X}\bs{M} + \bs{Y} \bs{N} &= \bs{I},\\
		\bs{M}^\bot\bs{M} + \bs{N}^\bot\bs{N} &= \bs{I}
	\end{split}
\end{equation} 
where $\bs{M}^\bot(s):=\bs{M}^\top(-s) \Rightarrow \bs{M}^\bot( \U{j}\omega )= \bs{M}^H(\U{j} \omega)$.
\end{dfn}
\begin{dfn}[NLCF]
Two matrices $\bs{\hat{M}} \in \mathcal{RH}_\infty^{p \times p}$ and $\bs{\hat{N}} \in \mathcal{RH}_\infty^{p \times q}$ form a normalized left coprime factorization (NLCF)
\begin{equation}
	\bs{G} = \bs{\hat{M}}^{-1} \bs{\hat{N}},
\end{equation}
if $\exists \bs{\hat{X}} \in \mathcal{RH}_\infty^{p \times p},\bs{\hat{Y}}  \in \mathcal{RH}_\infty^{q \times p}$ exists such that:
\begin{equation}
	\label{eq:NLCF}
	\begin{split}
		\bs{\hat{M}} \bs{\hat{X}} + \pmb{\hat{N}} \bs{\hat{Y}} &= \bs{I},\\
		\bs{\hat{M}}  \bs{\hat{M}}^\bot + \bs{\hat{N}} \bs{\hat{N}}^\bot &= \bs{I}.
	\end{split}
\end{equation}
\end{dfn}

The introduced normalized coprime factorizations can be used to define the following two graph symbols \cite[]{koenings2016}:
\begin{dfn}[NSIR]
	Let $\bs{G} = \bs{N} \bs{M}^{-1}$ be a NRCF of a system, then the normalized stable image representation (NSIR) of this system is defined as follows:
	\begin{equation}
		\bs{J} = \begin{pmatrix}
		\bs{M} \\ \bs{N}
		\end{pmatrix}
	\end{equation}
\end{dfn}
\begin{dfn}[NSKR]
	A normalized stable kernel representation (NSKR) of a system is defined by: 
	\begin{equation}
		\bs{K} = \begin{pmatrix}
		-\bs{\hat{N}} & \bs{\hat{M}}
		\end{pmatrix}
	\end{equation}
	where $\bs{G} = \bs{\hat{M}}^{-1}  \bs{\hat{N}}$ form the NLCF of this system.
\end{dfn}

Based on the before defined system representations the distance between systems can be defined.
\begin{dfn}
Let $\bs{K}_j$ be the NSKR of the system $\bs{G}_j$ and  $\bs{J}_i$ is the NSIR of the system $\bs{G}_i$. 
Then the $\nu$-gap-metric $d_{\nu} (\cdot,\cdot): \mathcal{R}^{p\times q} \times \mathcal{R}^{p\times q} \ra \mathbb{I}$ between the two systems $\bs{G}_i,\bs{G}_j$ is defined over the frequency domain $\forall \U{j} \omega \in (-\infty, \infty)$ as follows:
\begin{equation} \label{Eq:nugap}
	d_{\nu} (\bs{G}_i,\bs{G}_j) = 
	\begin{cases}
		\left\lVert\bs{K}_j\bs{J}_i \right\rVert_\infty&
		\begin{split}
			 &\det(\bs{K}^\bot_j\bs{J}_i)(\U{j} \omega) \neq 0 \\
			&\U{wno}\left(\det (\bs{K}_j^\bot\bs{J}_i )\right) = 0
		\end{split}\\
		1  &\text{otherwise}
	\end{cases}
\end{equation}
Where $\U{wno}\left(\det\left(\bs{G}\right)\right)$ denotes the winding number about the origin of $\bs{G},$ as $s\in \mathcal{D}$ follows the standard Nyquist D-contour $\mathcal{D}$ and 
\begin{equation*}
	\left\lVert \bs{K}_j\bs{J}_i \right\rVert_\infty \! = \!
	\left\lVert  (\pmb{I}  \!+ \! \bs{G}_j^\bot\bs{G}_j  )^{-\frac{1}{2}}
	 (\bs{G}_j  \! - \! \bs{G}_i  )
	 (\bs{I} \! + \! \bs{G}_i\bs{G}^\bot_i )^{-\frac{1}{2}} \right\rVert_\infty.
\end{equation*}
\end{dfn}

An intuitive interpretation is, that two systems can be stabilized by the same type of controller, if the $\nu$-gap metric between them is small.
A pointwise formula of the $\nu$-gap metric $\kappa$ is defined as follows
\begin{equation}\label{Eq:pointwise}
\kappa \left(\bs{G}_i (\U{j} \omega),\bs{G}_j (\U{j} \omega) \right) :=  \sigma_{\U{max}}(\bs{K}_j\bs{J}_i)(\U{j} \omega),
\end{equation}
where $\sigma_{\U{max}}(\cdot)$ is the maximum singular value.

%===============================================================================
% subsection stability margin
%===============================================================================
\subsection{Stability Margin and Maximum Stability Margin}
So far a distance between closed loop systems has been introduced, however a consistent way to design a controller for closely spaced systems, remains open.\\
For the design of controllers an essential relation between the $\nu$-gap metric and the stability margin $b_{\bs{G,}\bs{G}_\text{c}}$, can be exploited. 
\begin{dfn}[Stability Margin]
Let $\bs{G}\in \mathcal{R}^{p \times q}$ be a control system and $\bs{G}_\text{c}\in \mathcal{R}^{q \times p}$ a controller, then the stability margin $b_{\bs{G,}\bs{G}_\text{c}}$ is defined as:
\begin{equation}
	b_{\bs{G,}\bs{G}_\U{c}} :=
	\left\lVert \begin{pmatrix} \bs{I}\\ \bs{G}_\U{c} \end{pmatrix}
	(\bs{I}-\bs{G}_\U{c}  \bs{G})^{-1}
	\begin{pmatrix} \bs{I} & \bs{G} \end{pmatrix}
	\right\rVert_\infty^{-1}
	\label{eq:stability_margin}
\end{equation}
\end{dfn}
The stability margin $b_{\bs{G,}\bs{G}_\text{c}}$ and the $\nu$-gap metric are related according to the following theorem \cite{vinnicombe2001}:
\begin{thm}[Stabilizing Controllers]\label{thm:stabilising_c}
A controller $\bs{G}_\U{C}$ that stabilizes the system $\bs{G}_i$, also stabilizes the system $\bs{G}_j$ if and only if $d_\nu (\bs{G}_i,\bs{G}_j) < b_{\bs{G}_i,\bs{G}_\U{C}}$.
\end{thm}
Obviously the stability margin $b_{\bs{G}_i,\bs{G}_\U{C}}$ depends on the controller $\bs{G}_\text{c}$.
By taking into account all stabilizing controllers $\mathcal{G}_\U{c}$ for a specific plant $\bs{G}$ the maximum stability margin $b_{\text{max}}$ is defined as:
\begin{dfn}[Maximum Stability Margin]
Let \\ $||\bs{K}||_\text{H}$ be the \textsc{Hankel} norm of the NSKR of $\bs{G}$. 
Then 
\begin{equation}
	b_{\text{max}}=\sup\limits_{\text{stabilizing } \bs{G}_\U{c}}  b_{\bs{G,}\bs{G}_\U{c}}=
	\sqrt{1-||\bs{K}||^2_\text{H}}
\end{equation}
is called the maximum stability margin $b_{\text{max}}$ \cite[]{mcfarlane1992}.
\end{dfn}
Since any stabilizing $\bs{G}_\U{c}$ is considered $b_{\text{max}}$ becomes a system property, which can be exploited in the necessary and sufficient condition for robust stability of Theorem \ref{thm:stabilising_c}, see \cite{koenings2016}.
\begin{rem}\label{rem:b_max_d_nu}
	$b_\text{max}(\bs{G}_i)>d_\nu (\bs{G}_i,\bs{G}_j) \Leftrightarrow \exists \bs{G}_\U{C}$ that stabilizes the systems $\bs{G}_i$ and  $\bs{G}_j$.
\end{rem}
To actually design a controller with $b_{\bs{G}_i,\bs{G}_\U{C}}  \approx b_{\text{max}}$ we use a approach for robust stabilization presented in \cite{skogestad2001}.

%===============================================================================
% section Cluster Analysis and Construction of Prototypical Systems
%===============================================================================
\section{Cluster Analysis and Construction of Prototypical Systems}\label{sec:sys_aggregation}
The introduced metric $d_\nu$ forms a foundation to conduct a cluster analysis for a set of given systems which due to Remark \ref{rem:b_max_d_nu} %
%the relation of the maximum stability margin $b_\text{max}$ and $d_\nu$,%
will be based on the closed loop stabilization.
Moreover, the pointwise $\nu$-gap metric is used in an algorithm to find prototypical class descriptions for the cluster grouped systems.
%===============================================================================
% subsection cluster analysis
%===============================================================================
\subsection{Cluster Analysis} \label{sec:cluster_analysis}
Clustering is an unsupervised approach to partition a countable set of objects based on their similarity to discover data-inherent structures.\cite{}
In the the given problem the objects are simply transfer function matrices and the similarity is given by special case of the $\nu$-gap metric.
Based on the distance measurements between the systems $\mathcal{G}$ a cluster configuration $\mathcal{C}$ is sought, where every cluster 
$\mathcal{G}_i$ can be stabilized by one controller.
\\
For the cluster analysis no prior knowledge about the number and shape of the resulting clusters will be assumed. 
Furthermore to get reproducible cluster configurations a hierarchical cluster analysis, which is working monotonically by producing a set of nested clusters, will be applied.
Due to its monotony hierarchical cluster analysis can be visualized in form of a dendrogram which records the merging sequences of the clusters (see \reffig{fig:dendrogram}).
%\blue{
%This has another advantage, which you can split unstable classes ...
%}
\\
In order to get clusters which are expected to be stabilized with $b_\text{max}$, complete linkage as merging criterion was the natural choice, see \eqref{eq:complete_lnk}.
\begin{equation}\label{eq:complete_lnk}
	\begin{split}
	&d(\mathcal{G}_i,\mathcal{G}_j) = \max(d_\nu(\bs{G}_i,\bs{G}_j)) \\
	&\forall \bs{G}_i \in \mathcal{G}_i \text{ and } \forall \bs{G}_j \in \mathcal{G}_j
	\end{split}
\end{equation}
According to \eqref{eq:complete_lnk} clusters result in relation to the maximal distance between the systems and Remark \ref{rem:b_max_d_nu} can be exploited.

%===============================================================================
% subsection construction of new systems
%===============================================================================
\subsection{Construction of a System with Defined $\nu$-gap Metric} \label{sec:system_construction}
According to the problem statement, the next step comprises the aggregation of all the control systems in a cluster $\mathcal{G}_j$ into a prototype system $\bs{G}_{\text{proto},j}$, acting as class description for the controller design step. \\
Contrary to the rather arbitrary ``approach'' to choose a system $\bs{G}_i \in \mathcal{G}_j$ as a prototype, we provide an algorithm that constructs a prototype considering \emph{all} systems in the associated cluster $\mathcal{G}_j$\footnote{The resulting prototype $\bs{G}_{\text{proto},j}(s)$ will have a more general structure than $\forall \bs{G}_i \in \mathcal{G}_j$ }.
This section provides the formal basis for the constructions of systems with a predefined $\nu$-gap metric .
\\
Given a system $\bs{G}_1 $, it is possible to construct a new system $\bs{G}_2$ with an $\nu$-gap distance $d_{\nu} (\bs{G}_1,\bs{G}_2) <b_{\text{max}}$ using the following theorem \cite[]{vinnicombe2001}.  
\begin{thm} \label{thm:construction}
	Let $\bs{G}_1 \in \mathcal{R}^{p \times q}$, $\omega_\U{c} \in  \R$ and $\bs{H} \in \C^{p \times q}$. If
	\begin{equation}
		\beta = \kappa(\bs{G}_1(\U{j} \omega_\U{c}),\bs{H})<b_{\text{max}}(\bs{G}_1)
	\end{equation}
	there exists a  $\bs{G}_2 \in \mathcal{R}^{p \times q}$ satisfying
	\begin{equation}
		\bs{G}_2(\U{j} \omega_\U{c}) = \bs{H} \;\text{and } d_{\nu} (\bs{G}_1,\bs{G}_2) = \beta
	\end{equation}
\end{thm}
where $\kappa$ is the point-wise $\nu$-gap metric (\ref{Eq:pointwise}).

This theorem states that it is always possible to find a $\bs{G}_2$ whose $\nu$-gap distance to $\bs{G}_1$ is given by the upper bound $\beta$, which is obtained at the frequency $\omega_\U{c}$.
\\
The generation of a system $\bs{G}_2:d_{\nu} (\bs{G}_1,\bs{G}_2) = \beta$ is based on the construction of its NSIR $\bs{J}_2$ and follows the procedure below \cite[]{vinnicombe2001}.
Given the NSKR $\bs{K}_1$ and the NSIR $\bs{J}_1$ of the system $\bs{G}_1$ and let $\bs{\Omega} \in \mathcal{RH}_\infty^{p \times p}$ satisfy $\bs{\Omega}^\bot \bs{\Omega} = \bs{I}$ then 
\begin{equation} \label{Eq:J_2}
		\bs{J}_2 = \bs{J}_1 +  \bs{K}_1^\bot \bs{\Omega} \bs{\Delta}.
\end{equation}
In \eqref{Eq:J_2} $\bs{\Delta} \in \mathcal{RH}_\infty^{p \times q}$ has to satisfy the following conditions:
\begin{enumerate}[label=\alph*)]
	\item $\bs{\Delta} (\U{j} \omega_\U{c}) = \bs{\Omega}^\bot (\U{j} \omega_\U{c}) \bs{T}$
	\item $||\bs{\Delta}||_\infty = \sigma_{\U{max}}(\bs{T})$
	\item $\bs{\Delta}(\infty) = 0$ if $\omega_\U{c}<\infty$ or $\bs{\Delta}(0)=\bs{0}$ if $\omega_\U{c}=\infty$,
\end{enumerate}
where $\bs{T} \in \C^{p \times q}$ is defined as follows:
\begin{equation}
\bs{T} = \bs{K}_1 (\U{j} \omega_\U{c})  \begin{pmatrix} \bs{H} \\ \bs{I} \end{pmatrix}
\left( \bs{J}^\bot_1 (\U{j} \omega_\U{c})  \begin{pmatrix} \bs{H} \\ \bs{I} \end{pmatrix} \right)^{-1}.
\end{equation}
For the construction of such a \bs{\Delta}, first the singular value decomposition $\bs{\Omega}^\bot (\U{j} \omega_\U{c}) \bs{T} = \bs{T}_1 \Sigma \bs{T}_2$ is needed.
Then for $i=1,2$ $\bs{\Delta}_i$ is constructed as follows: 
\begin{equation*}
	\bs{\Delta}_i  \! = \! \left(\dfrac{s}{\omega_\U{c}} \Im \left\{ (\bs{T}_i \! - \! \bs{D})^{-1} \right\} \! + \! \Re \left\{ (\bs{T}_i \! - \! \bs{D})^{-1} \right\} \right) ^{-1} \! + \!  \bs{D},
\end{equation*}
where $\Re \left\{\cdot \right\}$ gives the real and $\Im \left\{\cdot  \right\}$ the imaginary part of a complex number and
$\bs{D}$ is chosen such that $\bs{D} \in \mathcal{R}^{p \times q}$ and $\bs{D}^\top\bs{D} = \bs{I}$.
To ensure that the resulting $\bs{\Delta} = \bs{\Delta}_1 \Sigma \bs{\Delta}_2$ is stable the unstable poles $p_k$ have to be reduced by multiplying every entry of the transfer function matrix $\bs{\Delta} $ with terms of the following form:
\begin{equation}
	F_{\U{P}k}(s) = \dfrac{(s-p_k)(s/\omega_\U{c}-\omega_\U{c}/p_k)}{(s + \bar{p}_k)(s/\omega_\U{c} - \omega_\U{c}/\bar{p}_k)}.
\end{equation}
Additionally it is required that the $\mathcal{H}_2$ norm of $\bs{\Delta}$ is finite. For this purpose $\bs{\Delta}$ will be multiplied by the term $\frac{\rho s}{s^2 + \rho s + \omega_\U{c}^2}$.
The parameter $\rho$ is chosen such that the conditions of Theorem \ref{thm:construction} are not violated.

%===============================================================================
% subsection prototype construction
%===============================================================================
\subsection{Construction of a Prototype for a Cluster of Systems } \label{sec:proto_alg}
This section addresses the construction of the prototypes for the clusters found by the hierarchical cluster analysis (see Section \ref{sec:cluster_analysis}). 
For the simplicity only systems with single input and single output are considered in the following, the MIMO case is treated analogously. 
\\
To preserve $\nu$-gap related properties a prototype is constructed in the way that it minimizes the maximal distance to the elements in its corresponding cluster. %
\begin{equation} \label{Eq:opt_problem}
	G_\U{proto} =  \argmin\limits_{G} \max\limits_{i}(d_{\nu} (G,G_{i}) ) \; \forall G_{i} \in \mathcal{G}_1
\end{equation}
The main idea for the construction is to change the prototype system $G_\U{proto}$ at the frequency $\omega_\U{c}$ which has the most influence on the $\nu$-gap metric between the prototype and the most distant system within a cluster.
For this purpose an iterative algorithm which adapts an initial prototype is presented in the following.
\\
The algorithm is initialized by choosing an initial prototype $G_\U{init}$ being the system with the smallest maximal distance to all the other systems in the cluster $\mathcal{G}_1$
\begin{equation}
	G_\U{init} =  \argmin\limits_{i} \max\limits_{j}(d_{\nu} (G_{i},G_{j}) ) \; \forall G_{i},G_{j} \in \mathcal{G}_1.
\end{equation}

Then the most distant system $\tilde{G}$ to the (initial) prototype is chosen and the ``worst case'' frequency $\omega_\U{c}$ is determined
\begin{equation}
	\omega_\U{c} = \argmax\limits_{\omega}\kappa \left(G_\U{proto}(\U{j} \omega),\tilde{G}(\U{j} \omega)\right).
\end{equation}
In the scalar case the pointwise $\nu$-gap metric $\kappa$ (\ref{Eq:pointwise}) simplifies to
\begin{equation} \label{Eq:pointwise_scalar}
	\kappa(G_{i}(\U{j}\omega),G_{j}(\U{j}\omega)) \! = \!
	\dfrac{ | G_{i}(\U{j} \omega) - G_{j}(\U{j}\omega) | }
 	{\sqrt{1 + |G_{i}(\U{j}\omega)|^2  } \sqrt{1 + |G_{j}(\U{j} \omega)|^2  } }.
\end{equation}

The next step realizes the adaption of the prototype at the ``worst case'' frequency $\omega_\U{c}$.
Therefor the optimization problem \eqref{Eq:opt_problem} is solved at $\omega_\U{c}$ 
\begin{equation}
	h = \argmin\limits_{G(\U{j}\omega_\U{c})} \max\limits_{i}\kappa (G(\U{j}\omega_\U{c}),G_i(\U{j} \omega_\U{c})) \; \forall G_i \in \mathcal{G}_1.
\end{equation}
If $\kappa(G_\U{proto}(\U{j} \omega_\U{c}),h) < b_{\text{max}}(G_\U{proto}(\U{j} \omega_\U{c}))$,
%Then Theorem \ref{thm:construction} is applied yielding a system $G_\U{proto}$ with $G_\U{proto}(\U{j}\omega_\U{c})=h$, which has the defined distance $\kappa (h,\tilde{G}(\U{j} \omega_\U{c}))=d_{\nu} (G_\U{proto},\tilde{G})$.
 Theorem \ref{thm:construction} can be applied yielding a system $G_\U{proto}$ with $G_\U{proto}(\U{j}\omega_\U{c})=h$, which has the defined distance $\kappa (h,\tilde{G}(\U{j} \omega_\U{c}))=d_{\nu} (G_\U{proto},\tilde{G})$.

To find such a system, the construction presented in Section \ref{sec:system_construction} is carried out where the parameter $\rho$ should be as high as possible, but  additionally as small as necessary.
For this reason we reduce $\rho$ step by step in a inner loop until we find a better system.
If no improvement occurs after a defined number $k_\text{max}$  of iterations, the algorithm terminates. 
Otherwise the prototype $G_\text{proto}$ is actualized. 
\\
The maximal pointwise $\nu$-gap metric may now be attained at another frequency $\omega_\U{c}$ leading to the reiteration of the algorithm.
If no better prototype can be found the previous one is take as final prototype $G_\U{proto}$.
\\
At the end the maximum stability margin $b_\U{max}$ of the prototype is calculated to check if all systems of the cluster can be stabilized by one controller.
%\footnote{ 
If this is not the case the cluster has to be split in smaller clusters with own prototypes.
This subclusters are already a result of the cluster analysis benefiting from the monotony of the hierarchic approach.\\
\blue{
% which is easy to implement without repeating the whole cluster analysis benefiting from the monotony of the hierarchic approach.} 
%
}
% , which means that the cluster analysis has to be repeated or reconfigured.}
The presented algorithm is summarized with the following  pseudocode, see below. 
\begin{algorithm}[!h]  \label{algh}
	input: $\mathcal{G}_1 = \{ G_1,\dots,G_\U{n} \}$ \\
	$G_\U{new} = G_\U{init} =  \argmin\limits_{i} \max\limits_{j}(d_{\nu} (G_{i},G_{j}) )$ \\
	\Repeat{$\max\limits_{i}d_{\nu} (G_\U{new},G_i ) - \max\limits_{i}d_{\nu} (G_\U{proto},G_i) > 0$}{
		$G_\U{proto} = G_\U{new}$\\
		$\tilde{G} = \argmax\limits_{G_{i}} d_{\nu} (G_\U{proto},G_{i})$ \\
		$\omega_\U{c} = \argmax\limits_{\omega}\kappa \left(G_\U{proto}(\U{j} \omega),\tilde{G}(\U{j} \omega)\right)$ \\
		$h = \argmin\limits_{G(\U{j}\omega_\U{c})} \max\limits_{i}\kappa (G(\U{j} \omega_\U{c}),G_i(\U{j} \omega_\U{c}))$ \\
		\For{$k = 1$  to $k_\U{max}$}{
			calculate new prototype $G_\U{new}(\U{j} \omega_\U{c})$ such that $G_\U{new}(\U{j} \omega_\U{c}) = h$ \\ 
			reduce $\rho$ \\
			\lIf{$\max\limits_{i}d_{\nu} (G_\U{new},G_i )< \max\limits_{i}d_{\nu} (G_\U{proto},G_i)$}
			 {$k = k_\U{max}$}
		}
	}
	check maximum stability margin: $b_\U{max}(G_\U{proto}) >  \max\limits_{i}d_{\nu} (G_\U{proto},G_i)$ \\
	output: $G_\U{proto}$
	\caption{Construction of a prototypical system for a cluster of similar systems}
\end{algorithm}
\vspace{-5mm}
%===============================================================================
% section Simulation/ Case Study
%===============================================================================
\section{Case Study}\label{sec:CaseStudy}
% \blue{MIMO-example necessary?}
In this section we apply our approach to a set of 80 systems including stable and unstable systems with proportional or integral behavior of different order.
The step responses of these different systems are shown in \reffig{fig:step_responses}. 
\subsubsection{Clustering.}	% Distance Measurement and Clustering.
\begin{figure}[!h]
	\centering
	\includegraphics[]{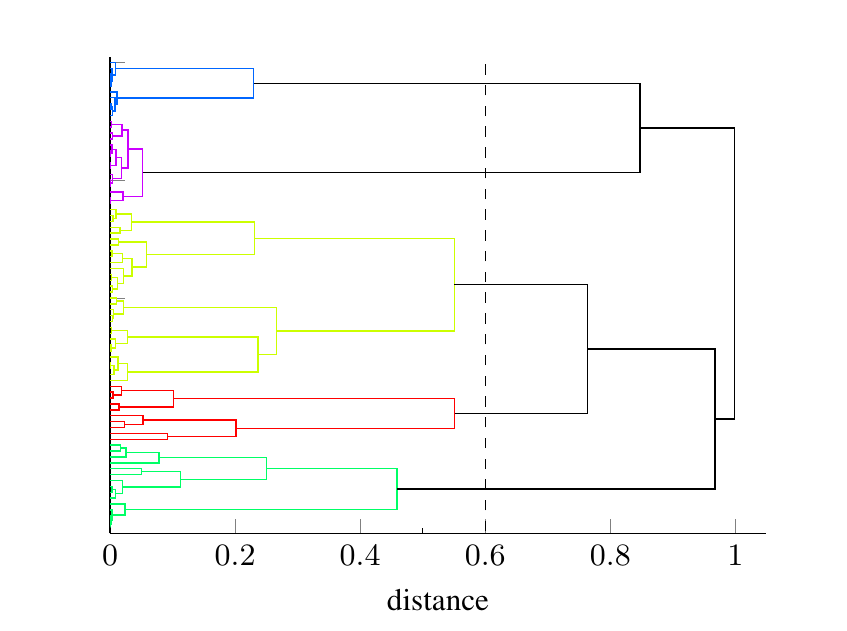}
	\caption{Complete linkage hierarchical clustering of the test set based on the $\nu$-gap distances.}
	\label{fig:dendrogram}
\end{figure}

\begin{figure}[!h]
	\centering
	\includegraphics[trim= 0 2.5mm 0 2.5mm]{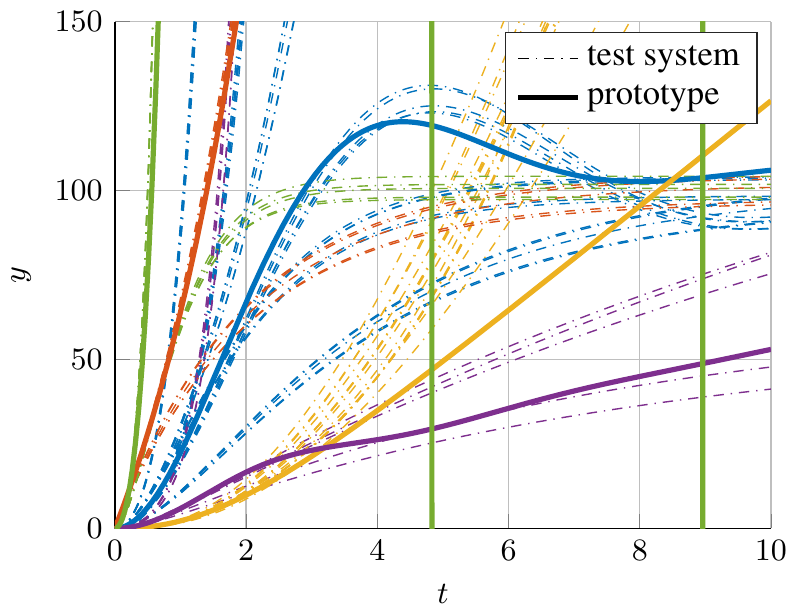}
	\caption{Step responses of the open loop systems.}
	\label{fig:step_responses}
\end{figure}

First the pairwise $\nu$-gap distances between the systems as well as their stability margins are calculated using the Matlab functions provides by \cite[]{auger2013}.
Then the systems are clustered as described in Section  \ref{sec:cluster_analysis}.
The outcome of the hierarchical cluster analysis using complete linkage is shown in the corresponding dendrogram (see \reffig{fig:dendrogram}).
% \blue{
% maybe example with cluster which can't be stabilized --- 
% cutting the dendrogram at a distance of $0.8$ the set of 80 systems is separated in four clusters 
% three stable ones and one: $\max d_{\nu} (G_{\U{proto},1},G_{i,1} )=0.49$,$b_\U{max}= 0.4$ $\ra$ split into two subclusters which can be stabilized ---
% use tree in the dendrogram  $\ra$ no new cluster analysis necessary  
% }
\\
Cutting the dendrogram at a distance of $0.6$ the set of 80 systems is separated in five clusters. 
Their actual cluster assignment is depicted in \reffig{fig:step_responses}.
It is noticeable that the open loop systems (see \reffig{fig:step_responses}) in a cluster are not necessarily similar.
This is because the systems are grouped with respect to their closed loop behavior.
\subsubsection{Prototype Construction.}
\begin{figure}[!h]
	\centering
	\includegraphics[trim= 0 2mm 0 0mm]{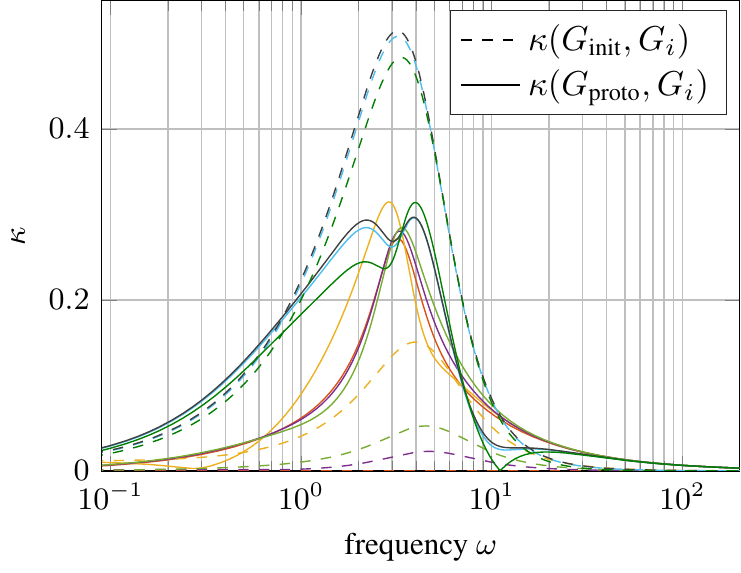}
	\caption{Pointwise $\nu$-gap metric between the initial and final prototype and the systems of cluster 1.}
	\label{fig:nu_gap_cluster4}
\end{figure}

Using Algorithm \ref{algh} the prototypes for each cluster are constructed.
Figure \ref{fig:nu_gap_cluster4} shows the first and last iteration of the algorithm for the prototype system $G_{\U{proto},1}$ of cluster 1.
Therein the dashed lines represent the pointwise $\nu$-gap metric between the initial prototype and all systems in the cluster $\mathcal{G}_1$.
Solid lines represent the pointwise  $\nu$-gap metric between the final prototype and all systems in the cluster.
After the applying of Algorithm \ref{algh} the maximal $\nu$-gap metric of the prototype $G_{\U{proto},1}$ to all $G_{i,1} \in \mathcal{G}_1$ could  be significantly improved, from $\max d_{\nu} (G_{\U{init},1},G_{i,1})=0.51$ at the start to $\max d_{\nu} (G_{\U{proto},1},G_{i,1} )=0.31$ at the end.  
\vspace{-4mm}
\subsubsection{Controller Design.}
\begin{figure}[!h]
	\centering
	\includegraphics[trim= 0 2mm 0 2mm]{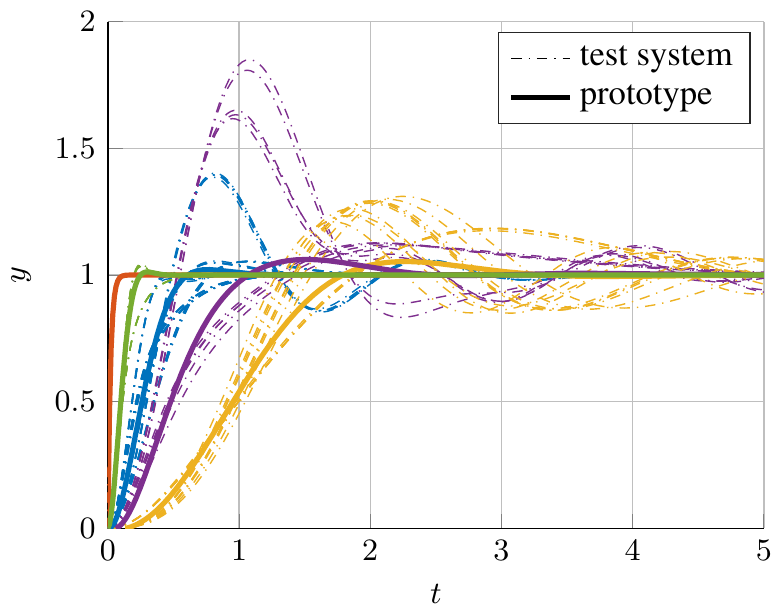}
	\caption{Step responses of the controlled systems.}
	\label{fig:cl_step_responses}
\end{figure}

In the next step the controllers, stabilizing all systems in a cluster, are constructed based on the prototypes.
To this end, the robust controller design presented by \cite{skogestad2001}, providing a stability margin as large as possible, will be applied.
Calculating the stability margins $b_{G,G_\U{C}} $ of the prototypes and the corresponding controllers using (\ref{eq:stability_margin}), we can guarantee the stability of the closed loop systems.
For the exemplary cluster the controller design based on $G_{\U{proto},1}$, results in a stability margin $ b_{G,G_\U{C}} = 0.40$.
Because the maximal distance, $ \max d_{\nu} (G_\U{proto},G_i) = 0.31$, is smaller than $ b_{G,G_\U{C}} $, this controller stabilizes all systems in the cluster 1.
\\
All remaining clusters have been treated with the same algorithm for the prototype and controller design.
The resulting closed loop system step responses for all clusters are summarized in  \reffig{fig:cl_step_responses}.
As depicted all systems in the set $\mathcal{G}$ could be stabilized by five different controllers.
However, if the $\nu$-gap between the prototype and their underlying system is high their dynamic behavior differ. %noticeable. 
\vspace{-5mm}
\subsubsection{Visualization.}
\begin{figure}[!h]
	\centering
	\includegraphics{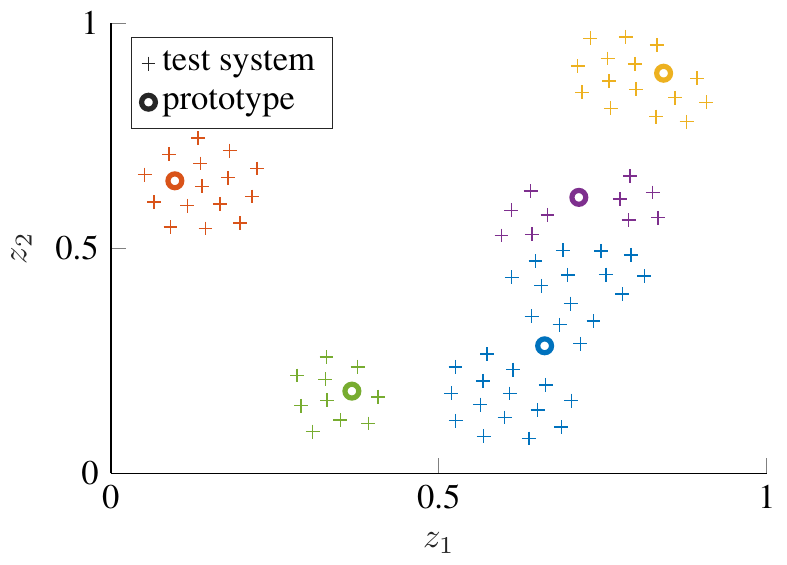}
	\caption{Visualization of the $\nu$-gap distances between the systems by  t-SNE mapping.}
	\label{fig:visualization}
\end{figure}

The systems of the case study are perceived as abstract object in an potentially high dimensional space---at least 85 dimensions.
For illustration purposes special dimension reduction mappings to get a two-dimensional representation of the cluster analysis and configuration can be applied. 
These mappings work in a way that every system $G_i$ is represented by a point  $\bs{z}_i= [z_1 \; z_2]^\top \in \R^2$ such that the respective metrics between the objects are preserved.\\ % $\bs{z}_i \in \R^2$
One possibility to find such a mapping is the so-called t-distributed Stochastic Neighbor Embedding (t-SNE) \cite[]{vandermaaten2008}.
In the first step the distances between the systems are converted into conditional probabilities 
\begin{equation}
	p_{i,j} = \dfrac{\e^{-d^2_{i,j} }} {\sum_{k \neq i} \e^{ -d^2_{i,k}  } } \; \forall i,j=1,\ldots, N.
\end{equation}
In the next step the so-called similarities $q_{i,j}$ for the low dimensional mapping are calculated by the use of a Student t-distribution
\begin{equation} %t-SNE
	q_{i,j} = \dfrac{ (1+|| \bs{z}_i-\bs{z}_j ||^2)^{-1}}{\sum_{k \neq i} (1 + ||\bs{z}_i-\bs{z}_k||^2)^{-1}  } \; \forall i,j=1,\ldots, N.
\end{equation}
To match this two distributions as well as possible, the Kullback-Leibler divergence given with the following cost function should be minimized
\begin{equation}
	J = \sum \limits_{i} \sum \limits_{j} p_{i,j} \log \dfrac{p_{i,j}}{q_{i,j}} \; \forall i,j=1,\ldots, N,.
\end{equation}
Figure \ref{fig:visualization} illustrates this mapping for the test set and the constructed prototypes.
The prototypes are the centers of the clusters with respect to the $\nu$-gap metric.
%\vspace{-2mm}
\vspace{-5mm}
	
%===============================================================================
% section Conclusion
%===============================================================================
\section{Conclusion and Outlook}
We have presented and adapted a machine learning methodology for the task of designing stabilizing controllers for sets of LTI systems. 
An important point of this approach was how to choose a suitable metric and technique of cluster analysis to partition a given set of linear systems into clusters for the controller design.
Due to the choice of the metric there are no restrictions regarding the model structures and orders. 
As a main contribution we derived an algorithm that constructs a prototype system for a given set of LTI systems.
Based on the prototype system we applied a method of controller design that guarantees the stability of the resulting closed loop for all systems assigned to the corresponding cluster.

\subsubsection{Outlook.} In our future work we like to use prototype systems as a classification model, since it aggregates all information of its underlying systems.
In an adaptive or hierarchical scenario, changing or new systems can be classified to the class/prototype with the shortest $\nu$-gap metric.
If this distance is shorter than the stability margin the system will be stabilized with the given controller.
Using an online estimation of the $\nu$-gap metric, as presented in \cite{koenings2018} or \cite{date2004}, the envisioned classification procedure of systems results in a switching procedure for the control, especially in combination with advanced classification concepts such as adaptive and multi-label classifiers and classification with rejection \cite[]{Hempel2013}.
To investigate the applicability  the approach have to be tested for multiple, various sets including  MIMO systems.
To improve our method the maximal stability margin should be included during the clustering and the prototype construction. 
% \blue{Take stability margin into account during clustering and prototype construction}
%Another interesting task is the description of nonlinear systems via a set of linear systems.
%Then a set of linear controllers can be automatically constructed using our approach.
\begin{ack}
We thank our reviewers for their thorough and valuable feedback.
\end{ack}
\bibliography{bib/Literatur_MACOF}             % bib file to produce the bibliography, with bibtex (preferred)
\end{document}